\documentstyle[12pt]   {article}
\def\dalemb#1#2{{\vbox{\hrule height .#2pt
        \hbox{\vrule width.#2pt height#1pt \kern#1pt
                \vrule width.#2pt}
        \hrule height.#2pt}}}

\let\a=\alpha \let\b=\beta \let\g=\gamma \let\d=\delta \let\e=\epsilon
\let\z=\zeta  \let\q=\theta \let\i=\iota \let\k=\kappa
\let\l=\lambda \let\m=\mu \let\n=\nu \let\x=\xi \let\p=\pi \let\r=\rho
\let\s=\sigma \let\t=\tau \let\u=\upsilon \let\f=\phi \let\c=\chi 
\let\w=\omega      \let\G=\Gamma \let\D=\Delta \let\Q=\Theta \let\L=\Lambda
\let\X=\Xi \let\P=\Pi  \let\U=\Upsilon \let\F=\Phi

 \def\bd{\begin{document}} \def\ed{\end{document}}
\def\ds{\documentstyle} \let\fr=\frac \let\bl=\bigl \let\br=\bigr
\let\Br=\Bigr \let\Bl=\Bigl 
\let\bm=\bibitem
\let\na=\nabla
\let\pa=\partial \let\ov=\overline
\def\ie{{\it i.e.\ }} 
\newcommand{\pr}{\paragraph{}}
\newcommand{\be}{\begin{equation}}
\newcommand{\ee}{\end{equation}}
\newcommand{\beba}{\begin{equation}\begin{array}{lcl}}
\newcommand{\eaee}{\end{array}\end{equation}}
\newcommand{\bea}{\begin{eqnarray}}
\newcommand{\eea}{\end{eqnarray}}
\newcommand{\ba}{\begin{array}}
\newcommand{\ea}{\end{array}}
\newcommand{\td}{\tilde}
\newcommand{\norsl}{\normalsize\sl}
\newcommand{\ns}{\normalsize}
\newcommand{\refs}[1]{(\ref{#1})}
\def\bal{{\mbox{\boldmath $\alpha$}}}
\def\bla{{\mbox{\boldmath $\lambda$}}}
\def\bbe{{\mbox{\boldmath $\beta$}}}
\def\bt{{\mbox{\boldmath $\tau$}}}
\def\bq{{\bf q}}
\def\bd{{\bf d}}
\def\bk{{\bf k}}
\def\bc{{\bf c}}
\def\bw{{\bf w}}
\def\bH{{\bf H}}
\def\bk{{\bf k}}
\def\bx{{\bf x}}
\def\boe{{\bf e}}
\def\a{\alpha}
\def\b{\beta}
\def\g{\gamma}
\def\c{\chi}
\def\d{\delta}
\def\e{\epsilon}
\def\ep{\varepsilon}
\def\FO{\phi}
\def\i{\iota}
\def\z{\psi}
\def\zb{\overline{\psi}}
\def\zt{\widetilde{\psi}}
\def\k{\kappa}
\def\l{\lambda}
\def\m{\mu}
\def\n{\nu}
\def\o{\omega}
\def\p{\pi}
\def\q{\theta}
\def\th{\theta}
\def\tc{\hat{\theta}}
\def\r{\rho}
\def\s{\sigma}
\def\st{\widetilde{\sigma}}
\def\sut{\utw{\sigma}}
\def\t{\tau}
\def\u{\upsilon}
\def\x{\xi}
\def\z{\zeta}
\def\w{\wedge}
\def\D{\Delta}
\def\F{\Phi}
\def\G{\Gamma}
\def\J{\Psi}
\def\L{\Lambda}
\def\O{\Omega}
\def\P{\Pi}
\def\Q{\Theta}
\def\U{\Upsilon}
\def\X{\Xi}
\def\f{\Phi_0}
\def\wtd{\widetilde}
\def\XH{\hat{X}}
\def\DH{\hat{D}}
\def\gh{\hat{g}}
\def\bh{\hat{b}}
\def\sg{\sqrt{-\g}}
\def\pa{\partial}
\def\gh{\hat{g}}
\def\bh{\hat{b}}
\def\mb{{\bar{m}}}
\def\nb{{\bar{n}}}
\def\rb{{\bar{r}}}
\def\sb{{\bar{s}}}
\def\ght{\hat{g}\kern-0.6em \widetilde{\raisebox{-0.12em}{\phantom{X}}}}
\def\bht{\hat{b}\kern-0.6em \widetilde{\raisebox{0.15em}{\phantom{X}}}}
\def\cF{{\cal F}}
\def\bG{{\bf G}}
\def\cL{{\cal L}}
\def\cM{{\cal M}}
\def\cG{{\cal G}}
\def\vf{\varphi}
\def\ft#1#2{{\textstyle{{\scriptstyle #1}\over {\scriptstyle #2}}}}
\def\fft#1#2{{#1 \over #2}}
\def\del{\partial}
\def\sst#1{{\scriptscriptstyle #1}}
\def\oneone{\rlap 1\mkern4mu{\rm l}}
\def\e7{E_{7(+7)}}
\def\td{\tilde}
\def\wtd{\widetilde}
\def\im{{\rm i}}
\def\bog{Bogomol'nyi\ }
\newcommand{\ho}[1]{$\, ^{#1}$}
\newcommand{\hoch}[1]{$\, ^{#1}$}
\newcommand{\ra}{\rightarrow}
\newcommand{\lra}{\longrightarrow}
\newcommand{\Lra}{\Leftrightarrow}
\newcommand{\ap}{\alpha^\prime}
\newcommand{\bp}{\tilde \beta^\prime}
\newcommand{\tr}{{\rm tr} }
\newcommand{\Tr}{{\rm Tr} } 

\begin{document}
\thispagestyle{empty}
\rightline{CTP-TAMU-20/98}
\rightline{hep-th/9805099}
\vspace{1truecm}

\centerline{\bf \Large Heterotic String Theory on Anti-de-Sitter Spaces}

\vspace{1.2truecm}
\centerline{\bf Karim Benakli\footnote{e-mail: 
 karim@chaos.tamu.edu}}
\vspace{.5truecm}
{\em
\centerline{ Center for Theoretical Physics,
Texas A\&M University, College Station, Texas 77843
}}

\vspace{2.2truecm}


\vspace{.5truecm}

\begin{abstract}

We suggest that compactifications on Anti-de-Sitter 
(AdS) spaces of type IIA, IIB, heterotic strings and eleven dimensional vacuua of M-theory are related by a combination of $T$ and strong/weak dualities. 
Maldacena conjecture relates then all these vacuua to a conformal theory on 
the boundaries. Furthermore acting with discrete groups on part of the 
internal spaces of these theories leads to further dual theories with less 
or no supersymmetry.

\end{abstract}

\let\LARGE=\large \let\Large=\large \let\large=\normalsize

\vfill {\small

}

\newpage


The last four years high energy physics has known a lot of excitement
as for the discovery of a set of dualities in certain class of
theories where quantum behavior is under control (supersymmetric and
finite theories)\cite{SEN}. These dualities state that models which seemed 
to be arising from totally different theories might describe the same physical 
system. Sometimes these dualities are symmetries, but often they are not but 
correspond to 
expansions around different values of the 
parameters of the theory. By studying these different limits, one hopes to 
reconstruct (and later understand) the full (non-perturbative) picture of 
the behavior of the system. In this short note we would like to consider 
dualities among different vacuua of string and M-theory on Anti-de-Sitter 
(AdS) spaces.  A detailed study of many related technical issues will be 
presented elsewhere \cite{tapr}.

Supersymmetric compactifications on AdS spaces often involve compact internal 
spheres. The cases of interest to us are those where either Type IIA or 
Type IIB superstrings internal space contains an $S^3$ factor. A well known
mathematical fact is that three spheres can be regarded as
Hopf fibration of $U(1)$ on $CP^1 \equiv S^2$
\be
 ds^2_{S^3}= ds^2_{S^2}+R^2 (dz+{\cal A})^2
\label{fibera}
\ee where ${\cal A}$ is a one-form on $CP^1$ and $R$ is the radius of
the sphere.  In contrast with supergravity theories that we believe
have a cut-off at the Planck scale, M-theory (and its stringy vacuua)
make sense even for very small radii. Following \cite{DLP}, one can
perform a $T$-duality on the $U(1)$ Hopf fiber. Because of the
non-trivial 3-forms appearing in the Type IIA and IIB compactifications
it was shown in \cite{DLP} that
$T$-duality maps $S^3$ vacuua of Type IIA (IIB)  to Type IIB (IIA) on 
$S^2 \times S^1$: 
\be 
ds^2_{S^3}= ds^2_{S^2}+\frac {1}{R^2} dz^2\; .
\label{fiberb}
\ee
The fiber is untwisted by the Hopf $T$-duality transformation. Combining this 
with strong/weak coupling duality will allow us to connect AdS vacuua of
type IIA, IIB, eleven dimensional M-theory and heterotic strings.

Our first example is Type IIB string theory on $AdS_3\times S^3\times
K3$.  Performing $T$-duality on the $S^3$ factor this theory is mapped
to IIA string theory on  $AdS_3\times S^2 \times S^1 \times K3$  where
now the radius of the $S^1$ factor is inversely proportional to the
one of the $AdS$ space.  Strong/weak coupling duality relates Type
IIA on  $K3$ to heterotic strings on $T^4$ or M-theory on $S^1 \times
K3$\cite{IIA}. We assume that the strong/weak coupling duality
between $IIA$ on $K3$ and heterotic on $T^4$ strings remains valid when
both spaces are  compactified to $S^2 \times S^1$ then  we can relate
the following set of vacuua\footnote{ I did not find reasons for a
possible failure of this duality although all the evidence for it
arises for cases with flat Minkowski spaces.}:

(1b) IIB string theory on $AdS_3\times S^3\times K3$,

(1a) IIA string theory on  $AdS_3\times S^2 \times S^1 \times K3$,

(1m) M-theory on $AdS_3\times S^2 \times T^2 \times K3$,

(1h) Heterotic on $AdS_3\times S^2 \times T^5$.

 Notice that although the $AdS$ radius is function of the string
coupling constant, it is  possible to choose the other parameters
large so that the volume of the  $AdS$ space remains large compare to the
string  scale in all these theories. This is necessary as it is
meaningless here to speak of $AdS$ geometry if the scales and
curvature are smaller than the Planck length\footnote {In the  case of
Type IIB on $AdS_5\times S^5 $ the $SL(2,Z)$ strong/weak coupling
symmetry is  preserved and the radius of the near horizon
Anti-de-Sitter space is kept large, this cooresponds in \cite{mald}
to have a large number of D3-branes.}.

The supergravity limit of model (1b) has attracted a lot of
interest recently \cite{BPS}. Starting from a configuration of
parallel $D5-$branes wrapped around $K^3$ and $D-$strings one can 
construct a self-dual string solution in six
dimensions \cite{DL}. Looking at the brane world-volume  theory and
taking the limit where  gravity decouples from the brane dynamics
leads to the large $N$ limit of  a conformal Yang-Mills theory in
$1+1$ dimensions. For the space-time geometry (as a solution of the
Type IIB equations of motion) the same limit gives rise  to Type IIB 
supergravity 
on $AdS_3\times S^3$. Following these observations, it was conjectured that 
supergravity on
$AdS_3\times S^3\times K3$ is dual to a (4,0) supersymmetric theory
in $1+1$ dimensions with an $SO(2,2)$ conformal symmetry
\cite{mald}. While evidence for the conjecture has been  exhibited only in 
the supergravity limit it is natural to extend it to the  level of the
full string theory. The precise map between the two theories follows
a holography principle and  has been explained in \cite{witten}.

The set of dualities presented above    involves the  exchange of
coupling constant on the Type IIA side with radius of the $AdS_3\times
S^2 \times S^1$ on the heterotic side, while the volume of the  $K3$ on 
the Type IIA side
determines the coupling constant on the  heterotic side. As a result
under these set of dualities $AdS_3\times S^2  \times T^5$
compactification  of heterotic strings would be related  to a (0,4)
superconformal theory in  $1+1$ whose Kac-Moody level is related to
the value of the heterotic  string coupling  ( as it is at the level of the 
worldsheet sigma model).

A similar way one can start with Type IIA dyonic string solution in six
dimensions \cite{DL}. The horizon geometry is given by  $AdS_3\times
S^3$ whose physics, following the holography principle, will be
encoded in a conformal field on the $1+1$ dimensional boundary.
Performing a Hopf T-duality on Type IIA on $AdS_3\times S^3   \times K3$
followed by strong/weak coupling dualities leads to relate the
following theories:

(2b) IIB string theory on  $AdS_3\times S^2 \times S^1 \times K3$,

(2a) IIA string theory on $AdS_3\times S^3\times K3$, 

(2m) M-theory on $AdS_3\times S^3 \times S^1 \times K3$,

(2h) Heterotic on $AdS_3\times S^3 \times T^4$.

The first of these theories (2b) can be regarded as the near-horizon
geometry of a  black string in five dimensions. Thus these
theories are also dual to a  superconformal theory in $1+1$
dimensions.

\vskip 0.3cm
Other possibilities for which we may apply the set of dualities are
compactifications of Type II strings on  $AdS_2 \times S^3$ spaces. 
The physics of
the $AdS_2$ space might be encoded in a quantum mechanical theory on
its one-dimensional boundary if the simplest version of the holography 
principle applies. 
This
supersymmetric quantum mechanics has an $SU(1,1)$ conformal symmetry and  has
been recently considered in \cite{AdS2}.  If the same set of dualities 
applies it will relate:

(3b) IIB string theory on  $AdS_2\times S^3 \times S^1 \times K3$,

(3a) IIA string theory on $AdS_2\times S^2 \times T^2 \times K3$, 

(3m) M-theory on $AdS_2\times S^2 \times T^3 \times K3$,

(3h) Heterotic on $AdS_2\times S^2 \times T^6$.

The latter theory is interesting because of the $SL(2,Z)$ strong/weak 
coupling symmetry of heterotic strings on $T^6$ \cite{tapr}. Exchanging 
the role of Type IIA and type IIB, we obtain the following set of  
dual compactifications:

(4b) IIB string theory on  $AdS_2\times S^2 \times T^2 \times K3$,

(4a) IIA string theory on $AdS_2\times S^3 \times S^1 \times K3$, 

(4m) M-theory on $AdS_2\times S^3 \times T^2 \times K3$,

(4h) Heterotic on $AdS_2\times S^3 \times T^5$.

\vskip 0.3cm
One can generate a set of other dual theories by ``dividing'' the
internal  spaces in the examples above by  their discrete subgroups
\cite {P}.  A particular simple  case of these  are freely acting
$Z_N$ on the $S^1$ factors. This $S^1$ might also be one of the $U(1)$
Hopf fibers \cite{DLP}. This generates a set of theories  with fewer
or no supersymmetries.  For example acting with a $Z_k$ on circle 
corresponds to keeping only states which have charges multiple of $k$ 
under the $U(1)$ isometries of the circle. For generic $k$ this leads to
to dualities between non-supersymmetric theories. This is due to the fact that
 the ``orbifolding'' does not 
act on the 
Anti-de-Sitter spaces, the theories are still dual to conformal theories 
on the boundaries with a corresponding number of supercharges \cite{kach}. 
Detailed results with many examples will be presented elsewhere \cite{tapr}.

If the dualities conjectured in this short note hold, they might
provide helpful to study the CFT/AdS dualities on the heterotic
side. These  involve tori and spheres with known metrics
instead of the complicate $K3$, some problems \cite{vafa} of the 
Type IIB side might hopefully be easier to investigate.

While composing the bibliography we came across  the work of
\cite{CVJ} where $AdS_3 \times S^3\times T^4$ of heterotic strings was
discussed in a different contest and with different results.

I wish to thank Michael Duff, Jian-Xin Lu, Chris Pope and Per Sundell
for  useful discussions.  This work was supported  by DOE grant
DE-FG03-95-ER-40917. 


\end{document}